\documentclass[pre,aps,superscriptaddress,showpacs,floatfix,notitlepage]{revtex4-1}
\usepackage{amssymb,mathrsfs,amsfonts}
\usepackage{amsmath}
\usepackage{graphicx}
\usepackage{epstopdf}
\usepackage{nicefrac}
\usepackage{listings}
\usepackage{upgreek}
\usepackage{multirow}
\usepackage{color}

\newcommand{\A}{{\cal A}}
\newcommand{\B}{{\cal B}}

\newcommand{\RN}[1]{%
  \textup{\uppercase\expandafter{\romannumeral#1}}%
}

\begin{document}
\title{Modeling of Supersonic Radiative Marshak waves using Simple Models and Advanced Simulations} 
\author{Avner P. Cohen}
\email{avnerco@gmail.com}
\affiliation{Department of Physics, Nuclear Research Center-Negev, P.O. Box 9001, Beer Sheva 84190, ISRAEL}
\author{Shay I. Heizler}
\email{highzlers@walla.co.il}
\affiliation{Department of Physics, Nuclear Research Center-Negev, P.O. Box 9001, Beer Sheva 84190, ISRAEL}


\begin{abstract}
We study the problem of radiative heat (Marshak) waves using advanced approximate approaches.
Supersonic radiative Marshak waves that are propagating into a material are radiation dominated (i.e. hydrodynamic motion is negligible), and can be described by the Boltzmann equation. However, the exact thermal radiative transfer problem is a nontrivial one, and there still exists a need for approximations that are simple to solve. The discontinuous asymptotic $P_1$ approximation, which is a combination of the asymptotic $P_1$ and the discontinuous asymptotic diffusion approximations, was tested in previous work via theoretical benchmarks. Here we analyze a fundamental and typical experiment of a supersonic Marshak wave propagation in a low-density $\mathrm{SiO_2}$ foam cylinder, embedded in gold walls. First, we offer a simple analytic model, that grasps the main effects dominating the physical system. We find the physics governing the system to be dominated by a simple, one-dimensional effect,  based on the careful observation of the different radiation temperatures that are involved in the problem. The model is completed with the main two-dimensional effect which is caused by the loss of energy to the gold walls. Second, we examine the validity of the discontinuous asymptotic $P_1$ approximation, comparing to exact simulations with good accuracy. 
Specifically, the heat front position as a function of the time is reproduced perfectly in compare to exact Boltzmann solutions. 
 
\end{abstract}

 \maketitle

\section{Introduction}
\label{s1}

Radiation heat waves play an important role in many high
energy density physics phenomena, for example in inertial confinement fusion
(ICF) and in astrophysical and laboratory plasma~\cite{Lindl2004}.
This problem has been a subject of both theoretical astrophysics research from early
times~\cite{Chandrasekhar1935,Milne1921}, and of experimental studies with high energy lasers facilities~\cite{lindl1995,Thomas2008}.
In recent decades, several experiments of supersonic (i.e. where the heat front velocity, is larger than the sound velocity) Marshak waves propagating through low-density foams were performed and reported. Typical published experiments are for example in~\cite{Massen,BackPRL,Back2000,Moore2015}. The theoretical understanding of these complicated systems is still incomplete~\cite{Fryer2016}.

The governing equation that describes the behavior of radiative heat waves is the radiative transport equation (RTE), also known as the Boltzmann equation (for photons). In the gray (``mono-energetic") radiation case the equation has this form:
\begin{equation}
\begin{split}
\frac{1}{c}\frac{\partial I(\hat{\Omega},\vec{r},t)}{\partial t}+\hat{\Omega}\cdot \vec{\nabla}I(\hat{\Omega},\vec{r},t)+
& \left(\sigma_{a}(T_m(\vec{r},t))+\sigma_{s}(T_m(\vec{r},t))\right)I(\hat{\Omega},\vec{r},t)=\\
& \sigma_{a}(T_m(\vec{r},t)){B}(T_m(\vec{r},t))+\\
& \frac{\sigma_{s}(T_m(\vec{r},t))}{4\pi}\int_{4\pi}I(\hat{\Omega},\vec{r},t)d\hat{\Omega}+
S(\hat{\Omega},\vec{r},t)
\end{split}
\label{Boltz}
\end{equation}
where $I(\hat{\Omega},\vec{r},t)$ is the specific intensity of radiation at position $\vec{r}$ propagating in the $\hat{\Omega}$ direction at time $t$. $B(T_m(\vec{r},t))$ is the thermal material energy, while the material temperature is $T_m(\vec{r},t)$, $c$ is the speed of light and $S(\hat{\Omega},\vec{r},t)$ is
an external radiation source. $\sigma_{a}(T_m(\vec{r},t))$ and $\sigma_{s}(T_m(\vec{r},t))$ are the absorption (opacity) and scattering cross-sections respectively.
Along with the equation for the radiation energy, the complementary equation for the material is:
\begin{equation}
\frac{C_v(T_m(\vec{r},t))}{c}\frac{\partial T_m(\vec{r},t)}{\partial t}=
\sigma_{a}(T_m(\vec{r},t))\left(\frac{1}{c}\int_{4\pi}{I(\hat{\Omega},\vec{r},t)d\hat{\Omega}}-aT_m^4(\vec{r},t)\right)
\label{Matter1}
\end{equation}
where $C_v(T_m(\vec{r},t))$ is the heat capacity of the material, and $a$ is the radiation constant ($a=4\sigma_\mathrm{SB}/c$, $\sigma_\mathrm{SB}$ is the Stefan-Boltzmann constant).

An exact solution for the transport equation is hard to obtain, especially in multi-dimensions. The most well-known exact approaches are the $P_N$ approximation, the $S_N$ method and Monte-Carlo techniques. In the $P_N$ approximation, we solve a set of moments equations when $I(\hat{\Omega},\vec{r},t)$ is decomposed into its first $N$ moments. The $S_N$ method solves the transport equation in $N$ discrete ordinates. These two approaches yield an exact solution of Eq.~\ref{Boltz} when $N\to\infty$~\cite{Pomraning1973}.
Alternatively, a statistically implicit Monte Carlo (IMC) approach can be used~\cite{IMC}. It is also exact when the number of particles (histories) goes to infinity.
Although these three methods tend to the exact solution, applying them requires massive simulation capabilities, and may be hard to solve, especially in multi-dimensions.
Therefore research for reliable approximations which will be easy to carry out, is important and useful.

Classic simple approximations, which are used in order to solve Eq.~\ref{Boltz}, such as the $P_1$ (as a simplest special case of the $P_N$ approximation) and diffusion approximations, suffer from several problems. 
They are both based on the assumption of isotropic (or close to isotropic) distribution of the specific intensity (which models correctly optically thick media)~\cite{Pomraning1973}. However, in a previous work~\cite{Cohen2018} we have derived a new time-dependent approximation, the {\em discontinuous asymptotic Telegrapher's equation approximation}, in a $P_1$ form. This approximation rests on two foundations: The asymptotic $P_1$ approximation~\cite{Heizler2010,Heizler2012,Ravetto_Heizler2012}, which leads to the correct asymptotic behavior and the correct front velocity,
and the discontinuous asymptotic diffusion of Zimmerman~\cite{zimmerman1979}, forcing a discontinuity of the energy density. The numerical solutions of this new approximation have been compared to well-known problems
such as the Su-Olson benchmark~\cite{SuOlson1996} and the Olson's nonlinear opacity problem~\cite{Olson2000}. It yields the best approximation for the exact results, comparing to other common approximations, including the gradient-dependent Flux-Limiter or Variable Eddington Factor approximations~\cite{Olson2000}. 

In this work we use the discontinuous asymptotic $P_1$ approximation that was described above, in order to analyze a fundamental supersonic Marshak wave experiment, carried out by C.A. Back et al.~\cite{Back2000}. In this experiment, a high energy laser is used to create a strong radiation heat source using a hohlraum, which then flows through a cylinder made of low-density $\mathrm{SiO_2}$ foam. This experiment was chosen since it is relatively optically thin. Hence, modeling the radiative transfer is challenging, and therefore interesting (simple diffusion approximation yields large errors). 
The experiment measures the out-coming flux from the foam cylinder in different foam lengths. Therefore the heat front position $x_F(t)$, can be extracted from the experimental results, in order to validate different theoretical models. We test the discontinuous asymptotic $P_1$ approximation via exact simulations (IMC and $S_N$), and various approximate methods. 

We start with a simple analytic model which yields a good understanding of the physical system. This includes one-dimensional (1D) prediction, and also considers two-dimensional (2D) effects. This work is an important verification of the new approximation in experimental circumstances, and it improves understanding of this experiment.  

The paper is structured as follows: First the classic diffusion and $P_1$ approximations are shortly introduced in Sec.~\ref{Approximate models for the Radiative Transfer Equation}. In Sec.~\ref{Mexperiment} we introduce the experimental configuration and details of the Back et al. experiment~\cite{Back2000}.
In Sec.~\ref{ExpDiff} an analytic model that offers a qualitative description of both main 1D and 2D effects of the problem is presented. In Sec.~\ref{Discontinuous} the discontinuous asymptotic $P_1$ approximation is introduced and tested. A short discussion is presented in Sec. \ref{discussion}.

\section{Classic approximation of the Boltzmann Equation}
\label{Approximate models for the Radiative Transfer Equation}

The first two angular moments of the specific intensity $I(\hat{\Omega},\vec{r},t)$ are:
\begin{subequations}
\label{EFdefine}
\begin{equation}
\label{Edefine}
E(\vec{r},t)=\frac{1}{c}\int_{4\pi}{I(\hat{\Omega},\vec{r},t)d\hat{\Omega}}
\end{equation}
\begin{equation}
\label{Fdefine}
\vec{F}(\vec{r},t)=\int_{4\pi}{ I(\hat{\Omega},\vec{r},t)\hat{\Omega}d\hat{\Omega}}
\end{equation}
\end{subequations}
where $E(\vec{r},t)$ is called the energy density, and $\vec{F}(\vec{r},t)$ is called the radiation flux.
Integration Eq.~\ref{Boltz} over all solid angles $\int{{d}\hat{\Omega}}$ yields the exact conservation law:
\begin{equation}
\\ \frac{1}{c}\frac{\partial E(\vec{r},t)}{\partial t}+
\\ \frac{1}{c} \nabla\cdot \vec{F}(\vec{r},t)=\sigma_{a}(T_m(\vec{r},t))\left(\int_{4\pi}{\frac{B(\vec{r},t)}{c}}d\hat{\Omega}-E(\vec{r},t)\right)+\frac{S(\vec{r},t)}{c}
\label{Rad1}
\end{equation}
Integration $\int{\hat{\Omega{d}}\hat{\Omega}}$ over Eq.~\ref{Boltz}, assuming that the specific intensity is a sum of its only two first moments (the well-known $P_1$ closure) yields:
\begin{equation} 
\frac{1}{c} \frac{\partial \vec{F}(\vec{r},t)}{\partial t}+\frac{c}{3}\vec{\nabla}E(\vec{r},t)+\sigma_{t}(T_m(\vec{r},t))\vec{F}(\vec{r},t)=0
\label{Rad2Class}
\end{equation}
when $\sigma_{t}(T_m(\vec{r},t))=\sigma_{a}(T_m(\vec{r},t))+\sigma_{s}(T_m(\vec{r},t))$ is the total cross-section.
The classic $P_1$ approximation contains the exact Eq.~\ref{Rad1}, along with the approximate Eq.~\ref{Rad2Class}. The classic diffusion (or the classic Eddington) approximation, assumes that the derivative of the radiation flux with respect to the time is negligible.
Therefore, Eq.~\ref{Rad2Class} yields a Fick's law form:
\begin{equation}
\vec{F}(\vec{r},t)=-\frac{c}{3\sigma_t(T_m(\vec{r},t))}\vec{\nabla}E(\vec{r},t),
\label{ficks}
\end{equation}

However, the classic diffusion approximation
yields an incorrect time-description because of its parabolic nature, i.e. an infinite particle velocity. The classic $P_1$ approximation (which can be presented as the hyperbolic Telegrapher's equation) yields an incorrect finite particle velocity~\cite{Heizler2010}.
The accuracy of the classic diffusion and $P_1$ approximations is tested against experimental data in the next sections. The diffusion approximation is also the base of the analytic model that will be presented in the Sec.~\ref{ExpDiff}.  

\section{Back et al. Experiment}
\label{Mexperiment}
During the last decades, experimental measurements of supersonic radiative heat
waves (Marshak waves) in low-density foams coupled to a high temperature source (i.e.~hot hohlraum), have been published ~\cite{Massen,BackPRL,Back2000,Moore2015}.
These experiments are part of the attempt to understand the macroscopic modeling of radiative hydrodynamics, which is crucial for high energy density physics (HEDP) and inertial confinement fusion (ICF).

\begin{figure}[htbp!]
\centering 
\includegraphics*[width=7.5cm]{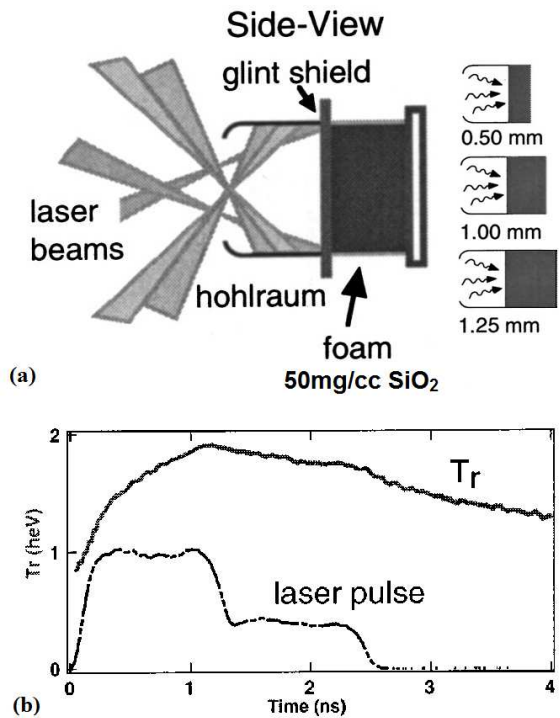}
\caption[BackExpSchem]{(a) Schematic diagram of the supersonic Marshak wave experiment by Back et al.~\cite{Back2000}. The primary hohlraum functions as a X-ray generator that flows into the low-density foam ($50\mathrm{mg/cm^3}$ $\mathrm{SiO_2}$), which is made in different lengths. X-ray steak camera measures the breakout radiation flux from the foam. (b) The measured hohlraum (drive) radiation temperature $T_D(t)$ as a function of the time in the Back experiment. The figure is taken from~\cite{Back2000}.}  
\label{fig:BackExpSchem}
\end{figure}

In these experiments, a powerful laser (energy$\approx$ 100J-1.8MJ with $t\approx$ 1-10nsec) irradiates a high-Z (usually gold) hohlraum, of about 1mm-1cm size. The energy is absorbed in the hohlraum's interior walls, and re-emits as soft X-rays with $T_{\mathrm{rad}}\approx$100-300eV as shown schematically in Fig.~\ref{fig:BackExpSchem}(a).

The experiment that is taken as a test-case is the experiment of Back et al. (see Fig.~\ref{fig:BackExpSchem})~\cite{Back2000}. This experiment was conducted in the Omega laser facility in Rochester. 10kJ of third-harmonic Nd:glass $0.35\mu$m laser with about $2.4$nsec long duration laser pulse heated a half hohlraum (halfraum).
The halfraum was converted the laser energy to a soft X-ray radiative drive that heats the $\mathrm{SiO_2}$ foam samples. The temperature drive $T_D(t)$ that was measured in the Back experiment is shown in Fig.~\ref{fig:BackExpSchem}(b). 
In this experiment they used samples of $50\mathrm{mg/cm^3}$ $\mathrm{SiO_2}$ of $0.5$, $1$, and $1.25$mm long, as presented in Fig.~\ref{fig:BackExpSchem}(a). The foam was coated by a $25\mu$m thick gold cylinder of $1.6$mm diameter.

We choose this experiment since $\mathrm{SiO_2}$ is a relatively optically thin material. Therefore, the difference between radiation models should be larger.
These experiments are also of sufficiently long duration, a factor which also increases the different between models. However, the impact of hydrodynamics in the foam can still be neglected, and the Marshak wave is supersonic (as opposed to the gold walls).

The primary diagnostics that was used in this experiment was an X-ray streak camera. It measures the flux that out-comes from the rear end of the sample as a function of time. Of special interest is the breakout time, which is when the flux is at approximately half-maximum intensity. Since the experiment is conducted with a different foam lengths, we can track the heat wave front, that is propagating through the sample.

\section{Simple analytic model for the calculating the heat front}
\label{ExpDiff}

In this section we introduce a simple model that is derived from the analytical solutions of one-dimensional radiative heat waves, both supersonic and subsonic~\cite{Marshak1958,Pakula1985,Zeldovich2002,rosenScale2,rosenScale3,HammerRosen,Shussman2015,Malka2016}. First, we introduce a one-dimensional (1D) model, which is based on the analytic solution for the supersonic Marshak waves of Hammer and Rosen (HR)~\cite{HammerRosen}, pointing out the importance of the different radiation temperatures that are involved on the problem~\cite{MordiLec,MordiPoster,Thomas2008}. Then, we expand the model to include the two-dimensional (2D) effect of the energy loss to the gold walls. This extension is based on the self-similar one-dimensional subsonic Marshak waves solutions for gold~\cite{Shussman2015,Malka2016}.

\subsection{The heat wave profile}

Radiative heat waves are characterized by a sharp front, due to the nonlinear behavior of the opacity and the heat capacity. There is a vast literature covering self-similar solutions of both supersonic and subsonic radiative (Marshak) heat-waves, assuming LTE conditions, i.e. $E(\vec{r},t)\approx aT_m^4(\vec{r},t)$~\cite{Marshak1958,Pakula1985,Zeldovich2002,rosenScale2,rosenScale3,HammerRosen,Shussman2015,Malka2016}. In these solutions, one assumes that the Rossland mean opacity $\kappa$ (which is connected to the absorption cross-section $\sigma_{a}(T_m(\vec{r},t))=\kappa\rho$, when $\rho$ is the material's density) and the internal energy $e(T,\rho)$ can be approximated in a power-law form (using~\cite{HammerRosen} notations):
\begin{subequations}
\begin{equation}
\frac{1}{\kappa}=gT^{\alpha}\rho^{-\lambda}
\label{opacityPowerLaw} 
\end{equation}
\begin{equation}
e=fT^{\beta}\rho^{-\mu}
\label{energyPowerLaw} 
\end{equation}
\end{subequations}
In Table~\ref{table:1} we introduce the different parameters for Silicon-dioxide ($\mathrm{SiO_2}$), which is the foam in the Back experiment and gold (the foam's walls). The parameters for $\mathrm{SiO_2}$ were fitted in the range of $100\mathrm{eV}\leqslant T\leqslant200\mathrm{eV}$, when for opacity (absorption) was fitted to a full CRSTA model tables~\cite{Kurz2012,Kurz2013}, and the equation-of-state (EOS) was fitted to QEOS model tables~\cite{QEOS}. According the CRSTA model the scattering coefficient in this temperature range is completely negligible. We note that the SESAME EOS table for silicon-dioxide~\cite{SESSAME} yields very similar parameters as the QEOS in these regimes. The parameters for gold were taken from~\cite{HammerRosen}.
\begin{table}
\begin{center}
\begin{tabular}{||c | c | c||} 
 \hline
 Parameter & $\mathrm{SiO_2}$ & Au \\[0.5ex] 
 \hline\hline
 $\alpha$  & $3.53$  & $1.5$  \\
\hline
$\beta$  & $1.1$  & $1.6$  \\
\hline
$\mu$  & $0.09$  & $0.14$  \\
\hline
$\lambda$  & $0.75$  & $0.2$  \\
\hline
$f$ [MJ]  & $8.78$  & $3.4$  \\
\hline
$g$ $\mathrm{[g/cm^2]}$ & $1/9175$  & $1/7200$  \\
\hline
\end{tabular}
\end{center}
\caption{The $\mathrm{SiO_2}$ and Au parameters that were used in this paper. The $\mathrm{SiO_2}$ parameters are fitted to exact opacity, CRSTA tables~\cite{Kurz2012,Kurz2013} and QEOS~\cite{QEOS} EOS tables in the $100\mathrm{eV}\leqslant T\leqslant200\mathrm{eV}$ regime. The parameters for Au were taken from~\cite{HammerRosen}.}
\label{table:1}
\end{table}

\begin{figure}[htbp!]
\centering 
\includegraphics*[width=7.5cm]{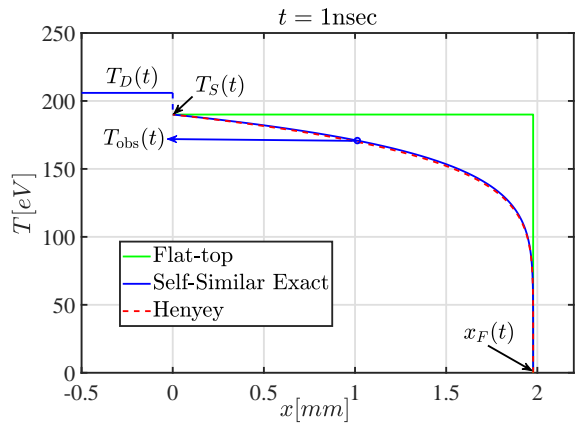}
\caption[Tprofile]{The temperature profile for a Marshak wave inside $\mathrm{SiO_2}$ under constant boundary conditions in 1nsec. The self-similar solution (blue) is compared to the approximated Henyey profile (red dashed). The three different radiation temperatures are marked, the drive (hohlraum) temperature, the surface temperature and the brightness temperature.}  
\label{fig:T_schem}
\end{figure}

Assuming that the boundary temperature also has a power law form, i.e. $T_S(t)=T_0t^{\uptau}$, a self similar solution can be achieved, both in the supersonic and the subsonic cases~\cite{Marshak1958,Pakula1985,rosenScale2,rosenScale3,Shussman2015,Malka2016}. The profile for constant temperature in 1nsec using an exact self-similar solution is shown in Fig.~\ref{fig:T_schem} (blue curves), for $T_S=190\mathrm{eV}$ in $\mathrm{SiO_2}$ with $\rho=50\mathrm{mg/cm^3}$ (close regime to the experiment). 
One can see the sharp profile which characterizes the Marshak wave (we can see in green a complete flat-top profile, $T(x,t)=T_S(t)H(1-x/x_F(t))$, when $H(x)$ is the Heaviside-step function). We note that for a specific power-law boundary condition, $\uptau=1/(4+\alpha-\beta)$, the temperature profile has a full analytic form, which is known as the Henyey solution~\cite{HammerRosen,MordiLec}:
\begin{equation}
T_\mathrm{Hy}(x,t)\approx T_S(t)\left(1-\frac{x}{x_F(t)}\right)^{\frac{1}{4+\alpha-\beta}}
   \label{Tshape} 
\end{equation}
$x_F(t)$ is the heat-front position, which is specific for $\uptau=1/(4+\alpha-\beta)$ Henyey case. We suggest, taking the Henyey profile and extending it for a general temporal behavior of the boundary condition, setting the correct $T_S(t)$ and $x_F(t)$ instead. For example, the temperature profile $T_\mathrm{Hy}(x,t)$ using the correct $x_F(t)$ for $\uptau=0$ (from the self-similar solution of~\cite{Shussman2015}), is shown
in red dashed curve in Fig.~\ref{fig:T_schem}. We can see the good accuracy of the ``Henyey-like" analytic profile. This profile will serve us subsequently, in the 2D-model.

\subsection{1D two (three) radiation temperatures model}

HR found a full analytic solution for 1D LTE supersonic diffusion equation using a perturbation expansion theory, for a general surface boundary condition $T_S(t)$~\cite{HammerRosen}. The heat front $x_F(t)$ is solved analytically and can be expressed as:
\begin{equation}
x_F^2(t)=\frac{2+\varepsilon}{1-\varepsilon}CH^{-\varepsilon}(t)\cdot\int_0^t H(t')dt',
\label{HRXF} 
\end{equation}
where:
\begin{subequations}
\begin{equation}
\varepsilon=\frac{\beta}{4+\alpha}
\end{equation}
\begin{equation}
C=\frac{16}{(4+\alpha)}\frac{g\sigma_\mathrm{SB}}{3f\rho^{2-\mu+\lambda}}
\end{equation}
\begin{equation}
H(t)=T_S^{4+\alpha}(t)
\end{equation}
\end{subequations}

\begin{figure}[htbp!]
\centering 
\includegraphics*[width=7.5cm]{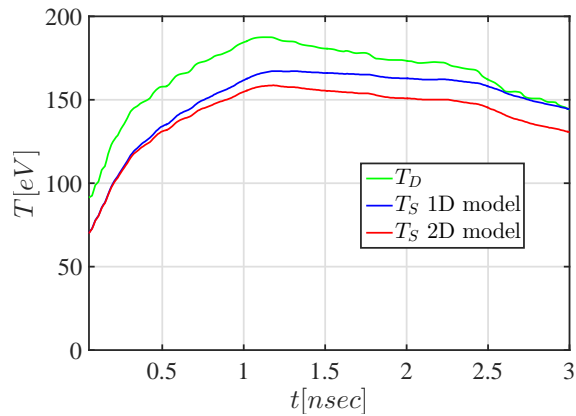}
\caption[Sprofile]{The drive temperature $T_D(t)$ (green) that was measured in the Back experiment, along with the surface temperatures $T_S(t)$ from the 1D model (blue) and the 2D model (red). The gap between $T_D(t)$ and $T_S(t)$ in the 1D model, is due to the re-emitted flux from the foam. The $T_S(t)$ in the 2D model is even lower because some of the energy is lost to the gold walls.}  
\label{fig:sourceProfile}
\end{figure}

\begin{figure}[htbp!]
\centering
\includegraphics*[width=7.5cm]{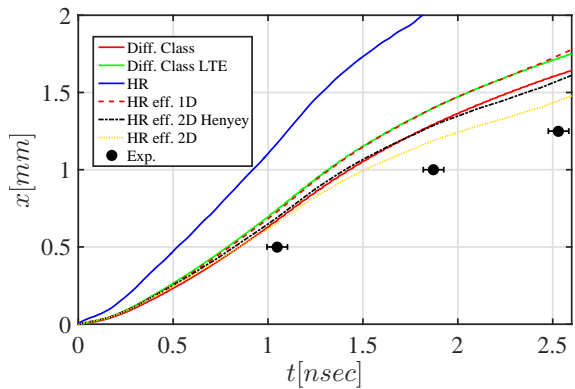}
\caption[XvsT]{The heat wave front $x_F(t)$ as a function of time for different simulations and models.
The experimental measurements are marked in circles and are taken from~\cite{Back2000}. The blue curve is the 1D {\em naive} HR model. Our 1D modification to HR is introduced in the red dashed curve, which fits to the LTE diffusion simulation (the green curve). Non-LTE diffusion simulation is in the solid red curve. The 2D model results are in the orange and black curves.}  
\label{fig:XvsT1}
\end{figure}

The key here is finding the surface temperature $T_S(t)$.~A {\em naive} assumption, that the surface temperature is equal to the radiation drive (hohlraum) temperature $T_D(t)$ (the green curve in Fig.~\ref{fig:sourceProfile}) for the Back experiment, yields a solution for $x_F(t)$ which is very far from the real experimental results. In Fig.~\ref{fig:XvsT1} the heat wave front $x_F(t)$ is presented as a function of time. It can be seen, that taking Eq.~\ref{HRXF} with $T_S(t)=T_D(t)$ (blue curve), yields heat front that is too fast, in fact twice faster than the experimental results (see also in Fig. 6 in~\cite{Back2000}). 
Other work modeled on the Pleiades experiments~\cite{Moore2015}, was also aware of that problem, and thus forced to apply an {\em ad hoc} multiplier on the surface temperature of $\approx0.71$ to fit the results.

Looking more carefully, one must distinguish three different radiation temperatures: The drive (hohlraum) temperature $T_D(t)$, the surface temperature $T_S(t)$ and $T_\mathrm{obs}(t)$, the brightness temperature that a detection will measure the re-emitted flux~\cite{MordiLec,MordiPoster}. The last is the temperature in $\approx1$mfp optical depth ($2/3$ mfp, assuming a LTE diffusion behavior)~\cite{MordiPoster}. To find the relation between the drive temperature and the surface temperature we apply a Marshak boundary condition~\cite{Olson2000,Thomas2008,MordiLec,MordiPoster}. Integrating the specific intensity over all incoming angles yields the incoming flux (assuming 1D slab geometry):
\begin{equation}
F_{D}(t)\equiv\sigma_\mathrm{SB}T_{D}^4(t)=\int_{-1}^0I(\mu,x,t)\mu d\mu
\end{equation}
Assuming the diffusion assumption for the specific intensity (a sum of its first two moments) yields (remembering that $E(0,t)\approx aT_S^4(t)$):
\begin{equation}
\sigma_\mathrm{SB} T_D^4(t)=\sigma_\mathrm{SB} T_S^4(t)+\frac{F(0,t)}{2}
\label{MarshakBC} 
\end{equation}
The relation between the drive temperature and the brightness temperature is due to energy conservation:
\begin{equation}
\sigma_\mathrm{SB} T_D^4(t)=\sigma_\mathrm{SB} T^4_{\mathrm{obs}}(t)+F(0,t)
\label{energy_cons} 
\end{equation}

Eqs.~\ref{MarshakBC} and~\ref{energy_cons} defines the relation between the three radiation temperatures. 
In Fig.~\ref{fig:T_schem} we can see the drive temperature $T_D$ assuming constant boundary condition $T_S=190\mathrm{eV}$ in $\mathrm{SiO_2}$ in 1nsec according Eq.~\ref{MarshakBC} using the self-similar solution for the flux $F(0,t)$, and the brightness temperature $T_{\mathrm{obs}}$ from Eq.~\ref{energy_cons} (or alternatively, the temperature in $2/3$ mfp). We can notice the non-negligible difference between the different radiation temperatures.

To apply the HR model for a general boundary condition $T_D(t)$ using Eq.~\ref{MarshakBC}, we need the complementary equation for $F(0,t)$. HR also denotes the energy that is stored in the material per unit area, recalling that $F(0,t)=\dot{E}(t)$:
\begin{equation}
E(t)=f\rho^{1-\mu}x_{F}(t)H^{\varepsilon}(t)(1-\varepsilon)
\label{EHR} 
\end{equation}

Now the 1D model is complete, solving Eqs.~\ref{HRXF},~\ref{MarshakBC} and~\ref{EHR}. Since the three integral equations are coupled, they can be solved easily, when a simple numerical algorithm is presented in the Appendix. Applying this simple model, only with the fidelity of the two different radiation temperatures, for the Back experiment, yields $T_S(t)$, as shown in Fig.~\ref{fig:sourceProfile} in the blue curve.
We can notice that there is a significant gap between $T_D(t)$ and $T_S(t)$. The complementary $x_F(t)$ for this model is shown in Fig.~\ref{fig:XvsT1} by the dashed red curve. 
We can notice the large improvement of the HR model (compared to the naive assumption), using the ``correct" $T_S(t)$, when the model yields $\approx1.3$ faster $x_F(t)$ than the experiment.

Testing the simple model, we apply a full 1D numerical radiation simulation of the same experiment, assuming LTE diffusion approximation (using the same power-laws for the opacity and the internal energy as in the model) with the given $T_D(t)$ (green curve in Fig.~\ref{fig:sourceProfile}). We can see the simulation's results in Fig.~\ref{fig:XvsT1} in the green curve, when the analytic solution and the simulation give the same results. 
The $\mathrm{SiO_2}$ foam that is used in this experiment is relatively optically thin material. Therefore, the LTE assumption is not accurate enough. Applying a full two-temperature (radiation and material) simulation, breaking the LTE assumption, yields the $x_F(t)$ that is shown in Fig.~\ref{fig:XvsT1} in the red curve. We can see that this simulation is a little bit closer to the experimental results than the LTE simulation/model. In Sec.~\ref{Discontinuous} we will test the models via more advanced numerical models, like IMC and $S_N$, and the new discontinuous asymptotic $P_1$, as well as flux-limiter diffusion approximations. 

We conclude this section by pointing out that the simple, analytic 1D model does identifies the basic physics and qualitative behavior of the system, however, with a disagreement with the experimental results. In order to address this, we offer a 2D model that includes the main 2D effect, the energy loss to the gold walls.

\subsection{2D (1.5D) effective model}

\begin{figure}[htbp!]
\centering 
\includegraphics*[width=14cm]{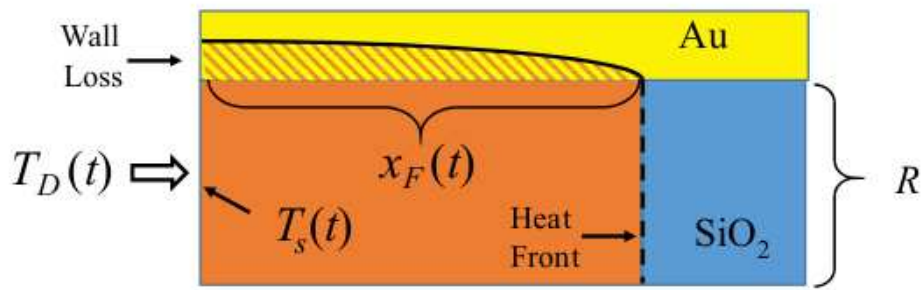}
\caption[WallLoss]{Schematic diagram of 2D slice of the heat wave in the new 2D model. The cold $\mathrm{SiO_2}$ and the gold wall are in blue and yellow, respectively (the gold area is larger than in reality). The orange area is the heated foam, while the yellow-orange pattern is the heated part of the gold. The heat wave loses energy to the gold walls, and therefore its velocity becomes slower. }  
\label{fig:WallLoss}
\end{figure}

For evaluating $x_F(t)$ that includes the energy loss to the walls, one can use the self-similar solutions of the 1D slab-geometry {\em subsonic} Marshak waves for gold~\cite{Shussman2015}. Schematic diagram for this 2D model (which we term the ``1.5 Model", since it is based on 1D modeling of the foam and 1D modeling of the gold) is shown in Fig.~\ref{fig:WallLoss}. The cold $\mathrm{SiO_2}$ and gold are shown in blue and yellow, respectively. The orange area is the heated area inside the foam, when the yellow-orange pattern is the heated area inside the gold walls. 

The simplest assumption is that the temperature profile inside the foam is flat-top, i.e. $T(x,t)=T_S(t)H(1-x/x_F(t))$ (the green curve in Fig.~\ref{fig:T_schem}). Thus, the gold surface temperature is $T_S(t)$ for all $0\leqslant x\leqslant x_F(t)$. For every space interval in $0\leqslant x\leqslant x_F(t)$, we set $t_0(x)$ as the time when the heat front reached that $x$.
The energy which is transferred toward the gold wall per unit area for time
$t$ can be taken from Shussman et al.~\cite{Shussman2015} (the gold parameters are taken from~\cite{HammerRosen} and are shown in Table~\ref{table:1}), when we choose the case of constant temperature boundary condition (since this matches more or less the $T_S(t)$ in the range of $1\mathrm{nsec}\leqslant t\leqslant 2.5\mathrm{nsec}$, see Fig.~\ref{fig:sourceProfile}):
\begin{equation}
E_{\mathrm{gold}}(t)=0.59T_0^{3.35}(t-t_0(x))^{0.59}  \quad\mathrm{[hJ/mm^2]}
   \label{Egold0} 
\end{equation}

Thus, the total energy $E_W(t)$ in the walls can be calculated by integrating Eq.~\ref{Egold0} with the area of the cylinder $2\pi Rdx$, when $R$ is the foam radius, considering the time duration that every space interval ``has seen" the heat wave:
\begin{equation}
E_W(t)=2\pi R \int_0^{x_F(t)}\int_{t_0(x)}^{t-t_0(x)}\dot{E}_{\mathrm{gold}}(t')dt'dx  \quad\mathrm{[hJ]}
   \label{EWall} 
\end{equation}

Now, the procedure is straightforward: The new model still solves Eqs.~\ref{HRXF},~\ref{MarshakBC} and~\ref{EHR} as in the 1D case, subtracting the incoming energy to the $\mathrm{SiO_2}$ foam (Eq.~\ref{EHR}) by the energy loss to the walls, Eq.~\ref{EWall}. Then, we re-estimate the new $T_S(t)$ and $x_F(t)$. Of course, this loss of energy decreases $T_s$, slows the heat front velocity. Thus, we call this ``1.5D" model, since it estimates the 2D effect by two 1D solutions, one for the foam and one for the gold.

\begin{figure}[htbp!]
\centering 
\includegraphics*[width=7cm]{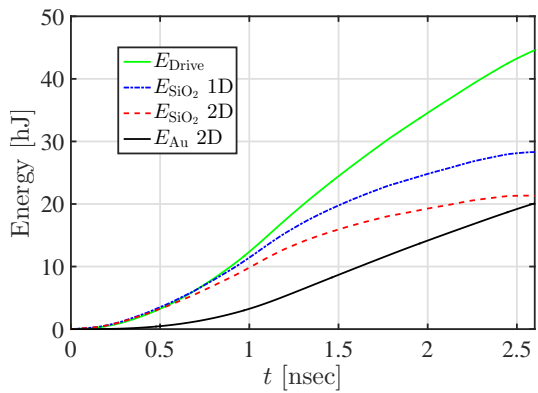}
\caption[Energy]{The different energies in the experiment. The incoming energy from the hohlraum ($E_D$) is in the green curve. The energy in the foam is 
presented by the dash-doted blue curve and the dash red line for the 1D and 2D model, respectively. The energy in the 2D model is lower because part of the energy is lost into the gold walls (black curve)}  
\label{fig:energies}
\end{figure}

The new surface temperature $T_s(t)$, using this 1.5D model is presented by the red curve in Fig.~\ref{fig:sourceProfile}, yielding lower $T_s(t)$, than the 1D model. The new estimation for the heat front $x_F(t)$ is shown in the orange curve in Fig.~\ref{fig:XvsT1}, yields slower heat front and much better results, showing that this 2D-effect is important. In Fig.~\ref{fig:energies} we can see the system energy in both the 1D and the 2D models. In the 1D model, some of the incoming energy ($E_D$, green line) heats the foam while some energy re-emits back into the hohlraum. However, in the 2D case, a significant amount of energy leaks to the gold walls (the black curve), decreases the amount of energy that is stored in the foam (the red dash line). Similar conclusions, that emphasis the importance of the 2D effects were also shown in~\cite{BackPRL,Back2000}, when simple estimations also reveal that a large amount of energy leaks to the walls, and the radiative heat wave propagation is no longer 1D.

One step further is to take the Henyey $T_\mathrm{Hy}(x,t)$ profile as in Fig~\ref{fig:T_schem} instead of the flat-top profile (that clearly, over-estimates the energy loss to the walls due to higher temperature near the front), into the integration of Eq.~\ref{EWall}. The black curve in Fig.~\ref{fig:XvsT1} shows the $x_F(t)$ results of this model, yields a faster results than the flat-top profile, due to lower energy loss to the walls, as expected. However, the difference is not large.

The remaining difference between the 2D model and the experiments is of course due to the assumption of LTE as indicated before, the transport model (that will be discussed in the next section) and due to more accurate opacity and EOS. In addition, another significant hydrodynamic 2D-effect is missing in this 2D-modeling, is the motion of the ablated gold (decreases $R$, the cylinder radius) especially near the foam surface, blocking part of the incoming energy from the hohlraum. Another 2D effect, is the spatial bending of the Marshak wave due to the wall~\cite{Hurricane2006}, but this effect is relatively small.

\section{The discontinuous asymptotic $P_1$ approximation}
\label{Discontinuous}

In a previous work, a modified $P_1$ approximation was offered, which is called the asymptotic $P_1$ approximation~\cite{Heizler2010,Heizler2012,Ravetto_Heizler2012}. In this approximation, a modified $P_1$ (that replaces Eq.~\ref{Rad2Class}) sets a modified equation, with two media-dependent coefficients, $\A(\vec{r},t)$ and $\B(\vec{r},t)$ (the $\A\B$ approximation):
\begin{equation} 
\frac{\A(\vec{r},t)}{c}\frac{\partial \vec{F}(\vec{r},t)}{\partial t}+c\vec{\nabla}E(\vec{r},t)+\B(\vec{r},t)\sigma_{t}(T_m(\vec{r},t))\vec{F}(\vec{r},t)=0
\label{Rad2H}
\end{equation}
where $\A(\omega_{\mathrm{eff}})$ sets the time-behavior and $\B(\omega_{\mathrm{eff}})$ reproduces the asymptotic diffusion coefficient~\cite{Case1953}. $\A(\omega_{\mathrm{eff}})$ and $\B(\omega_{\mathrm{eff}})$ are explicit functions of $\omega_{\mathrm{eff}}(\vec{r},t)$, the mean number of particles emitted per collision is defined as:
\begin{equation}
\omega_{\mathrm{eff}}(\vec{r},t)=\frac{\sigma_{s}E(\vec{r},t)+\sigma_{a}B(\vec{r},t)+S(\vec{r},t)/c}
{\sigma_{t}E(\vec{r},t)},
\label{omegaeff}
\end{equation}
Full explicit expressions for $\A(\omega_{\mathrm{eff}})$ and $\B(\omega_{\mathrm{eff}})$ are given in Appendix A in~\cite{Cohen2018},

When the asymptotic $P_1$ approximation is validated against exact solutions of known theoretical problems, it yields good results in homogeneous media~\cite{Heizler2010,Heizler2012}. This is due to the fact that far from sharp boundaries or strong sources, the asymptotic part is dominated.
Closer to the boundary, the transient part is more dominated and the continuous approximations yields poor behavior. To address this problem, previous works have solved the {\em exact} two-region semi-infinite half-spaces~\cite{mccormick1,mccormick2,mccormick3,ganapol_pomraning}, yielding solutions that have an asymptotic part that has a discontinuity in both the energy density $E$, and the energy flux $\vec{F}$ (and thus, do not conserves energy). 
Zimmerman~\cite{zimmerman1979} has derived a simple approximation for this two-region boundary problem. In this approximation, the first moment (the energy flux $\vec{F}(\vec{r},t)$) is continuous (and thus, the energy is conserved). However, the zero's moment (the energy density $E(\vec{r},t)$), is discontinuous, instead, $\mu(\vec{r},t)E(\vec{r},t)$ is continuous, where $\mu(\vec{r},t)$ is a function of the media properties~\cite{zimmerman1979,Pomraning1973,Cohen2018}. In order to derive a modified discontinuous diffusion approximation, Zimmerman expanded this method using a modified discontinuous Fick's law:
\begin{equation}
\vec{F}(\vec{r},t)=-\frac{cD(\vec{r},t)}{\mu(\vec{r},t)}\vec{\nabla}\left(\mu(\vec{r},t)E(\vec{r},t)\right),
\label{ZFick}
\end{equation}

The dependence of $\mu(\vec{r},t)$ in space and time is due to $\omega_{\mathrm{eff}}$, as $\A(\vec{r},t)$ and $\B(\vec{r},t)$. The full expressions for $\mu(\omega_{\mathrm{eff}})$ are also given in Appendix A in~\cite{Cohen2018}.
Following Zimmerman's rationale with a time-dependent analogy, and assuming that far from the boundary, the asymptotic $P_1$ is valid (Eq.~\ref{Rad2H}), and applying continuous flux and discontinuous energy density on the boundary, we yield a general discontinuous asymptotic $P_1$ equation:
\begin{equation}  
\mu(\vec{r},t)\frac{\A(\vec{r},t)}{c}\frac{\partial \vec{F}(\vec{r},t)}{\partial t}+c\vec{\nabla}\left({\mu(\vec{r},t)}E(\vec{r},t)\right)+
\mu(\vec{r,t})\B(\vec{r},t){\sigma_{t}((T_m(\vec{r},t))}\vec{F}(\vec{r},t)=0
\label{DisC2}
\end {equation}

Eqs.~\ref{Rad1} and~\ref{DisC2} define the new approximation, the {\em discontinuous asymptotic $P_1$ approximation}. These equations contain three medium-dependent variables, $\mu(\omega_{\mathrm{eff}})$ and $\A(\omega_{\mathrm{eff}})$ and $\B(\omega_{\mathrm{eff}})$, and thus we call it also the $\mu\A\B$ approximation. For more rigorous and detailed derivation, see~\cite{Cohen2018}.

\begin{figure}[htbp!]
\centering 
(a)
\includegraphics*[width=8.3cm]{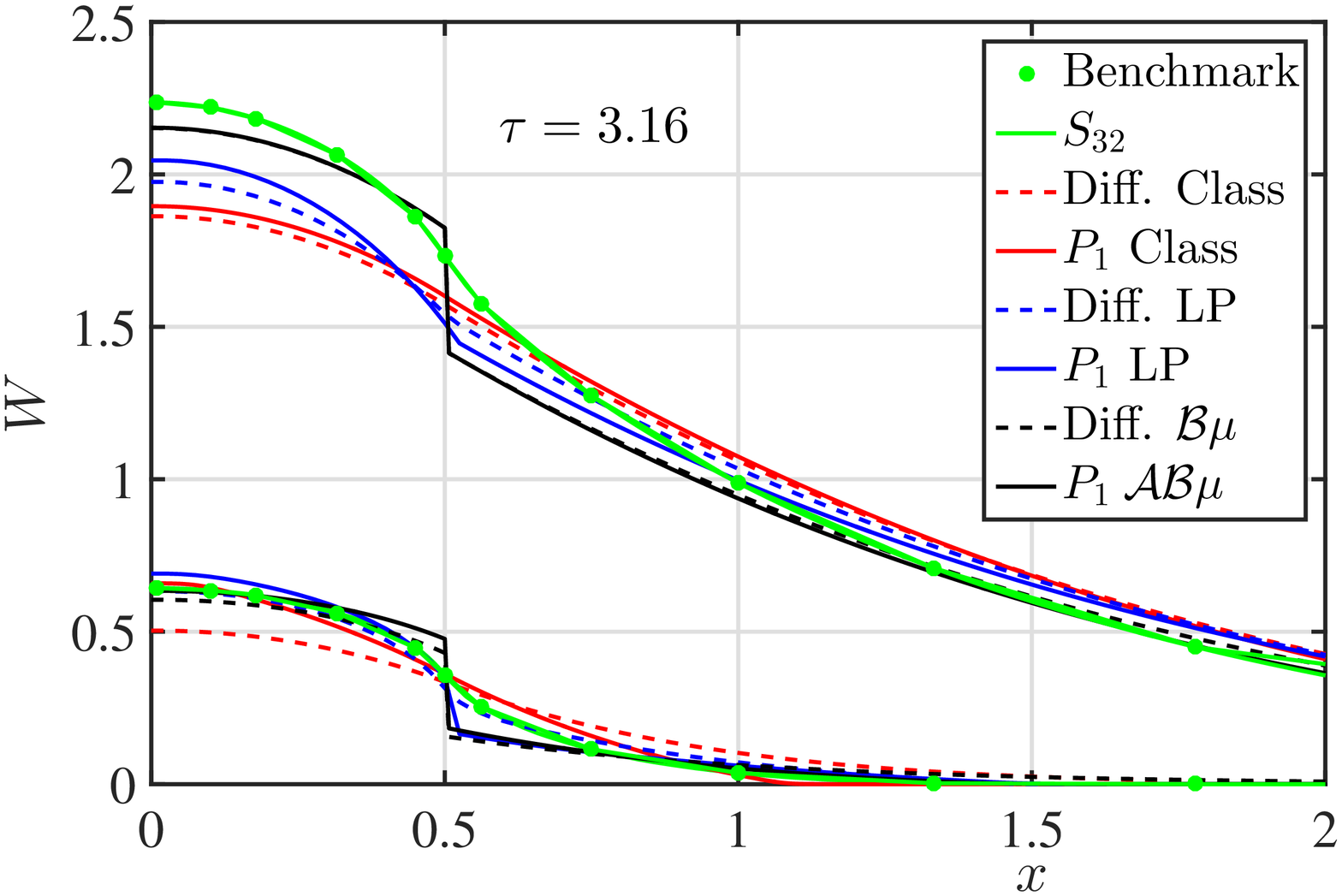}
(b)
\includegraphics*[width=8.3cm]{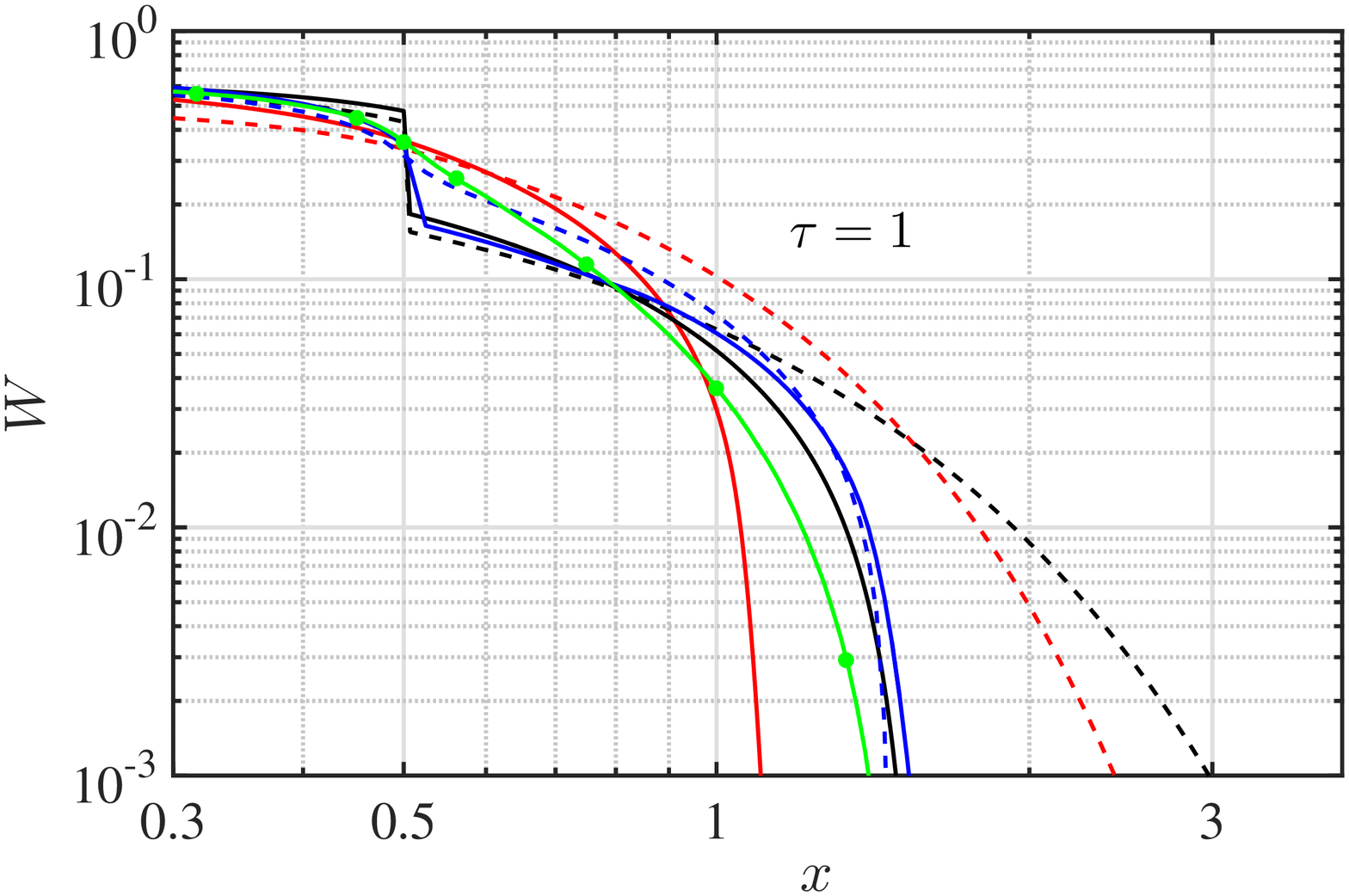}
(c)
\includegraphics*[width=8.3cm]{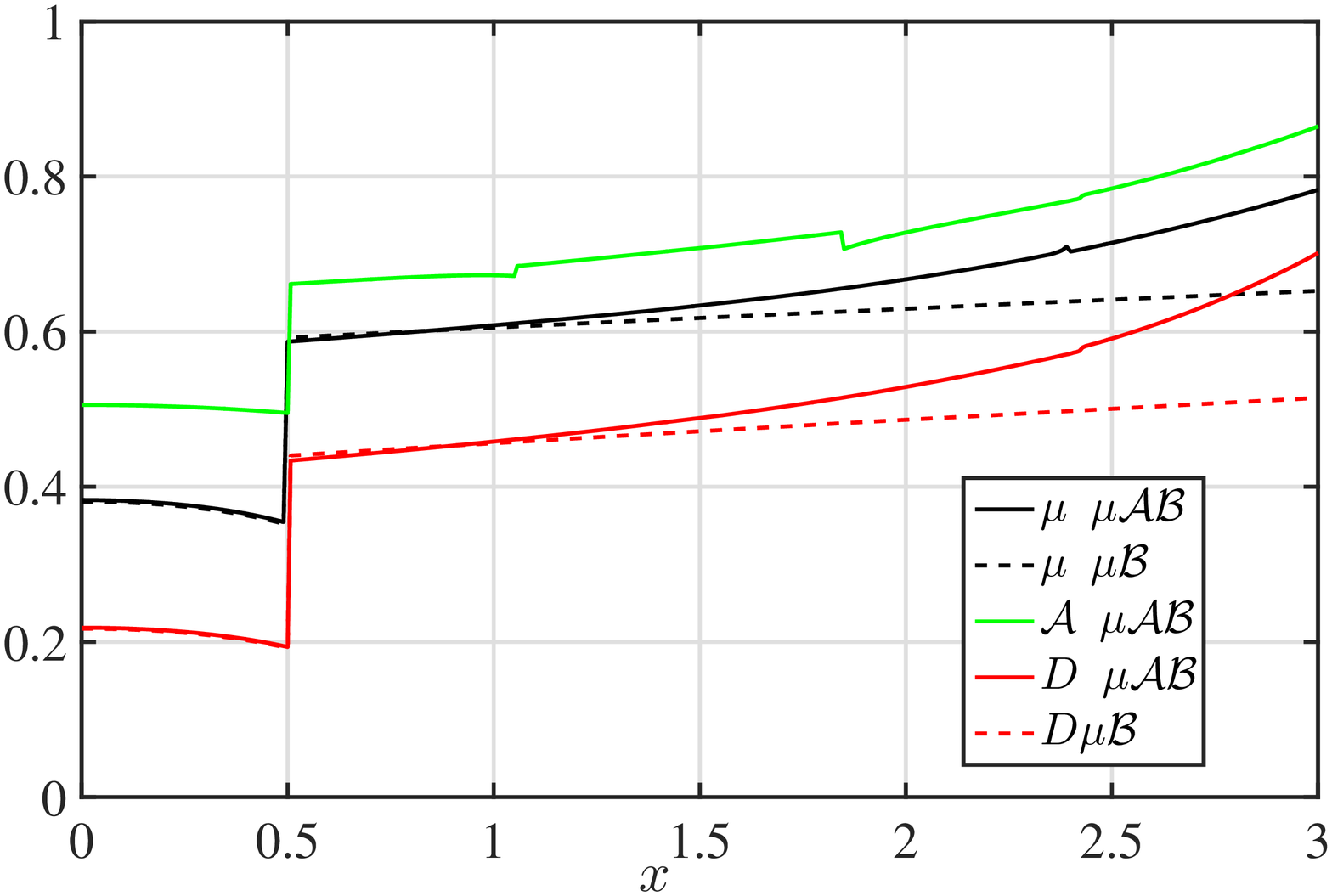}
\caption[SuOlson Problem basic]{The scaled radiation energy density (W) in linear (a) and logarithmic (b) scales as a function of the unitless optical depth, adopted from~\cite{Cohen2018}. The Su-Olson problem here is for pure absorbing case. The exact transport solution is in circles and taken from~\cite{SuOlson1996}. The green curves are the $S_{32}$ results. The red dashed and solid curves are the classic diffusion and $P_1$ approximations. The blue dashed and solid curves are the Levermore-Pomraning Flux-Limiter (dashed blue curves) and the Levermore-Pomraning Variable Eddington Factor (solid blue curves) results are presented. The black dashed and solid curves are Zimmerman's $\B\mu$ diffusion and the $\A\B\mu$ $P_1$ approximations. (c) The coefficients $\A(\omega_{\mathrm{eff}})$, $\B(\omega_{\mathrm{eff}})$ and $\mu(\omega_{\mathrm{eff}})$ as a function of the space in $\tau=3.16$ for $\mu\B$ and $\mu\A\B$ approximations.}
\label{fig:SuOlsonBasic}
\end{figure}

The new discontinuous asymptotic $P_1$ approximation was tested numerically, with the well-known constant opacity Su-Olson benchmark~\cite{SuOlson1996}. Both the Su-Olson benchmark and the Back experiment are optically thin, pure-absorbing problems. For example, the scaled radiation energy as a function of space ($W=\int_{-1}^{1}d\mu\frac{I(\mu)}{aT_H^4}$, when $T_H$ is the Hohlraum reference temperature), is presented in Fig.~\ref{fig:SuOlsonBasic} for the pure absorbing case.
In Fig.~\ref{fig:SuOlsonBasic}(a) the radiation energy is shown as a function of $x$, the scaled spatial variable in linear scale for $\tau\equiv c\sigma_{t}t=$3.16, 10 and in Fig.~\ref{fig:SuOlsonBasic}(b) in logarithmic scale for $\tau=1$. In Fig.~\ref{fig:SuOlsonBasic}(c) we plot the dimensionless parameters $\A(\omega_{\mathrm{eff}})$, $\B(\omega_{\mathrm{eff}})$ and $\mu(\omega_{\mathrm{eff}})$ as a function of the $x$ for $\tau=3.16$ for the $\mu\A\B$ and Zimmerman's $\mu\B$ approximations. We can see the discontinuities in the different parameters in the boundary of the source, when both $\B$ and $\mu$ are very similar in both approximations, except the front, when Zimmerman's approximation yields too fast heat front; in Zimmerman's approximation $\A=0$, of course.

As it can be seen that both the classic diffusion and $P_1$ approximations yields too low bulk energy (Fig.~\ref{fig:SuOlsonBasic}(a)). In addition, in the logarithmic scale (Fig.~\ref{fig:SuOlsonBasic}(b)) it is noticeable that the heat front of the classic diffusion approximation propagates too fast, while the classic $P_1$ is too slow. We note that the front of the asymptotic $P_1$ is quite good but its bulk energy is too small and similar to the classic $P_1$ approximation~\cite{Cohen2018}. Zimmerman's discontinuous asymptotic diffusion approximation (the $\mu\B$ approximation), yields better results in the bulk, which is resulting from the discontinuity jump condition, but the heat front is still too fast. However, it is clear that the new discontinuous asymptotic $P_1$ approximation (the $\mu\A\B$ approximation) is very close to the exact solution, both in the bulk and in the front. It is even better than the gradient-dependent approximations, both the Flux-Limiters and Variable Eddington Factors results~\cite{Olson2000}.

In the previous section, we saw that the standard diffusion approximation (even the non-LTE version) yields non-accurate results. That is because $\mathrm{SiO_2}$ is relatively optically thin material, so this problem is challenging to diffusive models. We will now test our new model, versus exact transport models such IMC and $S_N$, in the Back experiment. The test here is limited to 1D calculations, so 2D-effects are not included in this section. Thus, the comparison of the different models is to the exact transport models (IMC and $S_N$), and not directly to the experiments.

\begin{figure}[htbp!]
\centering 
(a)
\includegraphics*[width=7.5cm]{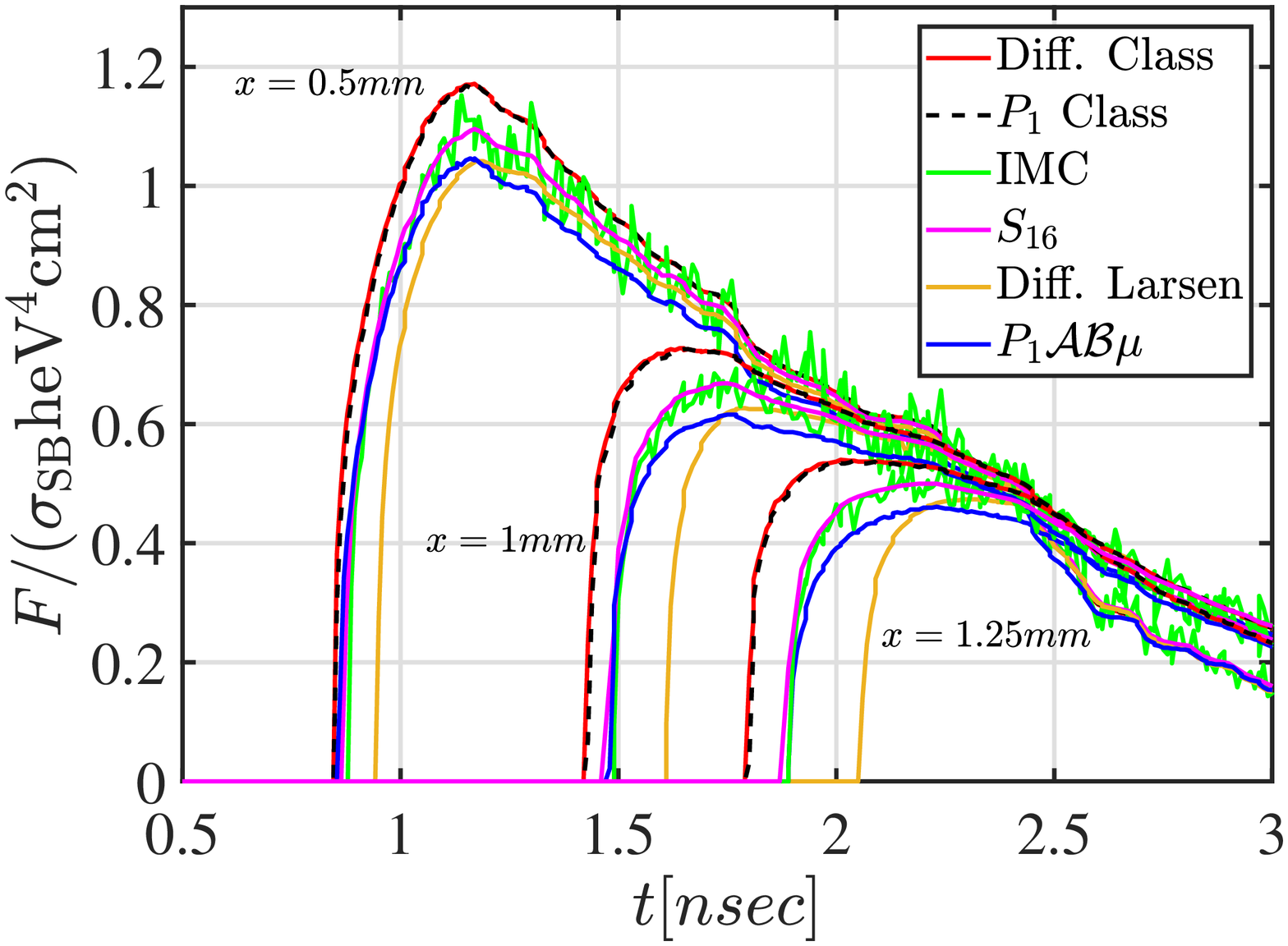}
(b)
\includegraphics*[width=7.5cm]{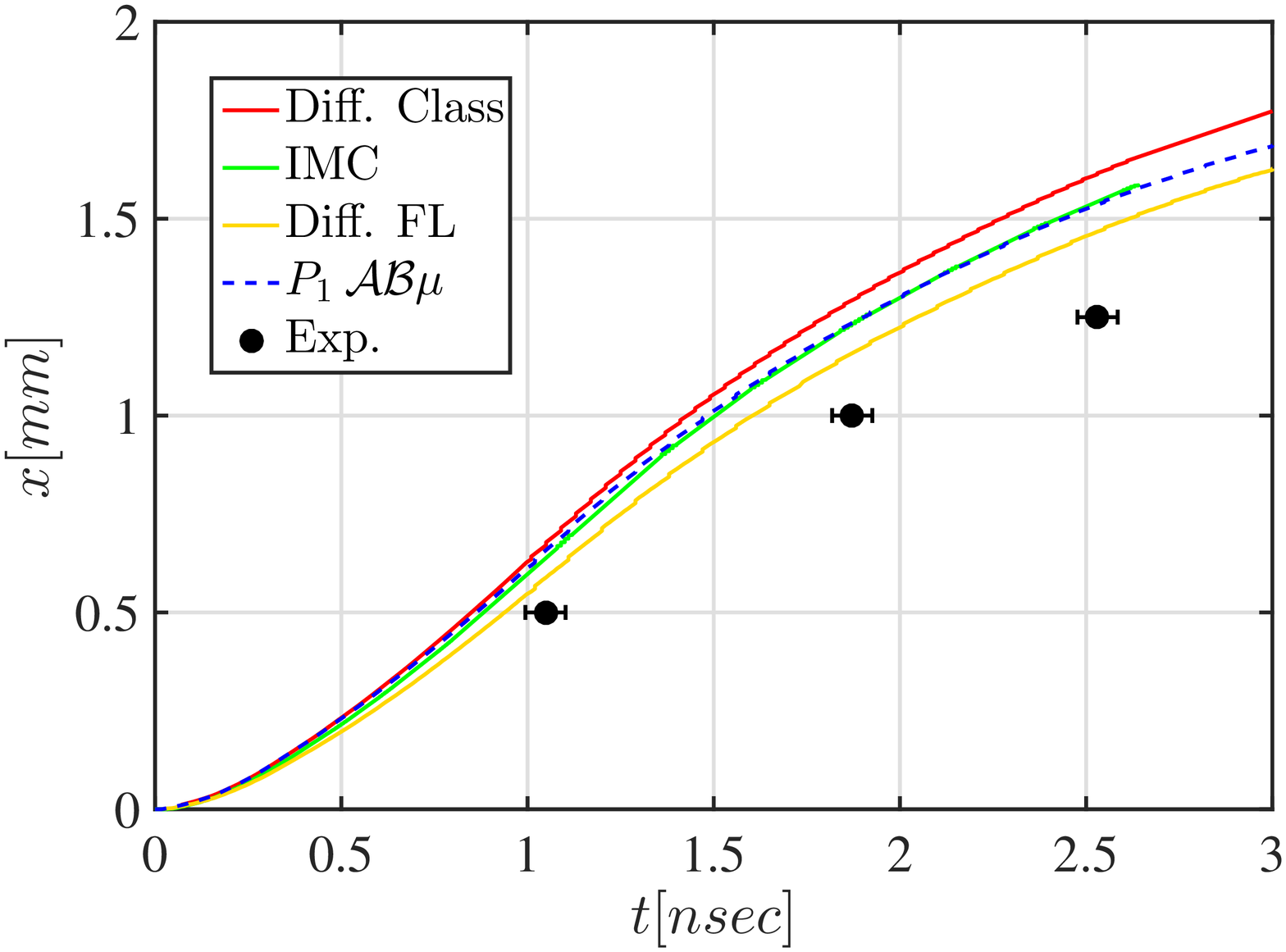}
\caption[Fprofile]{(a) The out-coming flux from the end of the foam cylinder as a function of time for the different lengths of foam using different transport models. It can be seen than both diffusion and $P_1$ are faster than the exact IMC and $S_N$ models, while the Flux-limited diffusion is slower. The new discontinuous $\mu\A\B$ approximation yields similar results to the exact ones. (b) The heat front position as a function of time using different transport models. Again, the $\mu\A\B$ approximation yields very good agreement with the exact models.}  
\label{fig:Fprofile}
\end{figure}

First, we apply both 1D slab-geometry IMC and $S_N$ using the $T_D(t)$ of the Back experiment. Next, we used various of approximate transport models, including standard diffusion and $P_1$, flux-limited diffusion and finally, the new discontinuous asymptotic $P_1$ ($\mu\A\B$) approximation. It is important to note that all the approximate diffusive and $P_1$ approximations used the exact boundary condition~\cite{Olson2000}:
\begin{equation}
F(0,t)=2F_{D}(t)-\frac{c}{2}E(0,t)
   \label{MarshakZBC} 
\end{equation}
Eq.~\ref{MarshakZBC} is a non-LTE generalization of Eq.~\ref{MarshakBC}.
The results of the out-coming flux for the different foam lengths are presented in Fig.~\ref{fig:Fprofile}(a), while the heat front position $x_F(t)$ is presented in Fig.~\ref{fig:Fprofile}(b), for the different transport models.

We can see in Fig.~\ref{fig:Fprofile}(a) that exact IMC and $S_N$ codes yield precisely the same results, and are considered to be accurate in 1D. The non-LTE diffusion simulation yields faster heat-front than does the exact simulations (Fig.~\ref{fig:Fprofile}(b)) and higher flux for all the foam's lengths (Fig.~\ref{fig:Fprofile}(a)). The classic $P_1$ yields the same results as the classic diffusion and shows no improvement. The Flux-Limited diffusion (using Larsen's FL with $n=2$~\cite{Olson2000}) yields $x_F(t)$ that is slower than the exact simulations. However, the new discontinuous asymptotic $P_1$ ($\A\B\mu$) approximation yields very close results, comparing to the IMC and $S_N$ results.
The fluxes of the $P_1$ $\mu\A\B$ approximation are a little bit lower than the exact simulations, but finds the exact breakout times. Overall, the discontinuous asymptotic $P_1$ approximation yields the best approximate solution comparing the exact approaches. Since we have seen in Sec.~\ref{ExpDiff} the 2D-effects are important to simulate the Back experiment, we plan to extend the discontinuous asymptotic $P_1$ approximation to 2D, to find out if it yields closer results to the experiment.

Finally we note that using an asymptotic-distribution based boundary condition:
\begin{equation}
F(0,t)=2F_{D}(t)-\mu(\omega_{\mathrm{eff}})cE(0,t)
   \label{MarshakZBC2} 
\end{equation}
yields worth results than using Eq.~\ref{MarshakZBC}, thus we recommend using Eq.~\ref{MarshakZBC} instead of Eq.~\ref{MarshakZBC2}.

\section{Discussion}
\label{discussion}
The solution of the radiative transfer problem which is coupled to the material under realistic conditions is complicated. We present in our paper several methods solving the Boltzmann equation and tested them in two different problems. Table~\ref{table:2} summarizes the different simulations which were compared. Table~\ref{table:3} summarizes the different {\em analytic} models which were used.

We have here managed to describe a major phenomenon by a simple analytic model and approximate numerical simulations. The understanding of the difference between the drive temperature that is measured/calculated inside the primary hohlraum ($T_D(t)$) and the material boundary temperature ($T_S(t)$), is crucial for this analysis. This model which is based on the HR solution shows good results with the experiments. In this simple model the Marshak boundary condition and the energy conservation define the relationship between the drive temperature ($T_D(t)$) and the boundary temperature ($T_S(t)$) that were used. In previous works~\cite{BackPRL,Back2000,Moore2015} the gap between $T_D(t)$ and $T_S(t)$ was not marked. This distinction is important for better modeling of these kind of experiments.
The 2D model produces even better results, showing the major 2D effect caused by the energy loss to the gold walls. It is important to note that we limit all the models in this work to a power-laws approximate opacity and EOS. 

\begin{table}
\begin{center}
\begin{tabular}{||l | l| c | l||} 
 \hline
  & Method  & In Figures  & Basic assumptions  \\[0.1ex] 
 \hline\hline
1 & IMC Simulation & \ref{fig:Fprofile} &  Statistical implicit Monte-Carlo approach.\\
\hline
2 & \multirow{2}{*}{$S_N$ Simulation} & \multirow{2}{*}{\ref{fig:SuOlsonBasic}, \ref{fig:Fprofile} } & Solves the transport equation in $N$ \\& & & discrete ordinates.\\
\hline
3 & Classic Diffusion  & \multirow{5}{*}{\ref{fig:XvsT1}}  & The specific intensity is a sum of its only two first\\ &LTE  Simulation & &moments ($\int_{4\pi}{ I(\hat{\Omega},\vec{r},t)\hat{\Omega}\hat{\Omega}d\Omega}\approx E(\vec{r},t)/3$),\\  & & &the derivative of the energy flux $\vec{F}(\vec{r},t)$ with \\  & && respect to time inside Eq.~\ref{Rad2Class} is negligible,\\  & & &and LTE conditions, ($E(\vec{r},t)\approx aT_m^4(\vec{r},t)$).
  \\
\hline
4 & Classic Diffusion  & \multirow{2}{*}{\ref{fig:XvsT1}, \ref{fig:SuOlsonBasic}, \ref{fig:Fprofile} } & The same as 3\\ & Non-LTE  Simulation& & without LTE conditions.  \\
\hline
5 & \multirow{2}{*}{Classic $P_1$  Simulation} & \multirow{2}{*}{\ref{fig:SuOlsonBasic}, \ref{fig:Fprofile}}  & The specific intensity is a sum of its only two
first \\& & &moments ($\int_{4\pi}{ I(\hat{\Omega},\vec{r},t)\hat{\Omega}\hat{\Omega}d\Omega}\approx E(\vec{r},t)/3$).  \\
\hline
6& LP (Levermore-Pomraning)& \ref{fig:SuOlsonBasic}  & General diffusion approximation\\&Flux-limiter Simulation && when the diffusion coefficient is, $D(\vec{r},t)=\frac{\lambda(R(\vec{r},t))}{\omega_{\mathrm{eff}}(\vec{r},t)}$,\\& & & $\lambda(R(\vec{r},t))=\left[\coth(R(\vec{r},t))-\frac{1}{R(\vec{r},t)}\right]\frac{1}{R(\vec{r},t)}$,\\& & &  and $R(\vec{r},t)=\frac{\vert\vec{\nabla}{E(\vec{r},t)} \vert}{\omega_\mathrm{eff}(\vec{r},t)\sigma_{t}(T_m(\vec{r},t))E(\vec{r},t)}$  \\
\hline
7 &Larsen's ($n=2$)  & \ref{fig:Fprofile}  & General diffusion approximation  \\
&Flux-limiter Simulation& & when the diffusion coefficient is,\\& & &$D(\vec{r},t)=\left[(3\sigma_{t}(T_m(\vec{r},t)))^n+\left(\frac{1}{E(\vec{r},t)}\frac{\partial{E(\vec{r},t)}}{{\partial{x}}}\right)^{n}\right]^{-\nicefrac{1}{n}}$\\
\hline
8 &LP (Levermore-Pomraning)& \ref{fig:SuOlsonBasic}   & General $P_1$ approximation when: \\
&Eddington-factor Simulation  & &$\int_{4\pi}{ I(\hat{\Omega},\vec{r},t)\hat{\Omega}\hat{\Omega}d\Omega}=\vec{\nabla}(\chi(\vec{r},t)E(\vec{r},t))$ \\
& & & $\vec{f}(\vec{r},t)$, the ratio between the first two moments:
\\
& & & $\vec{f}(\vec{r},t) = \frac{\vec{F}(\vec{r},t)}{cE(\vec{r},t))}$, when
\\
& & & 
 $\vert\vec{f}(\vec{r},t)\vert=\coth(z(\vec{r},t))-1/z(\vec{r},t)$ and\\
& & & $ \chi(\vec{r},t)=\coth(z(\vec{r},t))[\coth(z(\vec{r},t))-1/z(\vec{r},t)]$.\\

\hline
9 &Discontinuous asymptotic & \ref{fig:SuOlsonBasic}  & Asymptotic diffusion approximation,  \\
&diffusion  Simulation & & $\vec{F}(\vec{r},t)$ is continuous and $\mu_AE_A(\vec{r},t)=\mu_BE_B(\vec{r},t)$\\
\hline
10 &Discontinuous asymptotic & \ref{fig:SuOlsonBasic}, \ref{fig:Fprofile}  & Asymptotic $P_1$ approximation,  \\
&$P_1$ Simulation& &$\vec{F}(\vec{r},t)$ is continuous and $\mu_AE_A(\vec{r},t)=\mu_BE_B(\vec{r},t)$\\
\hline
\end{tabular}
\caption{All the simulations were used in our paper, with their basic assumptions}
\label{table:2}
\end{center}
\end{table}

\begin{table}
\begin{center}
\begin{tabular}{||l | l| c | l||} 
 \hline
  & Method  & In Figures  & Basic assumptions  \\[0.5ex] 
 \hline\hline
11 &HR 1D Naive & \ref{fig:XvsT1}  & Self-similar solution of the LTE diffusion equation, \\
&Analytic Model  & & when $T_S=T_D$.\\
\hline
12 &HR 1D & \ref{fig:sourceProfile}, \ref{fig:XvsT1}, \ref{fig:energies} & Self-similar solution of the LTE diffusion equation, \\
&Analytic Model  & & when $T_S$ is found from the Marshak boundary conditions.\\
\hline
13 &HR 2D & \ref{fig:XvsT1}  & Self-similar solution of the LTE diffusion equation, \\
&Analytic Model - Henyey & & when $T_S$ is found from the Marshak boundary conditions\\ & & & and from the energy leakage into the gold walls.\\
& & & The heat wave is taken as ``Henyey-like" shape.\\
\hline
14 &HR 2D & \ref{fig:XvsT1}  & Self-similar solution of the LTE diffusion equation, \\
&Analytic Model& & when $T_S$ is found from the Marshak boundary conditions\\ & & & and from the energy leakage into the gold walls.\\
& & & The heat wave is taken as flat-top shape.\\
\hline
\end{tabular}
\end{center}
\caption{All the analytic methods were compared in our paper, with a short explanation}
\label{table:3}
\end{table}

In the Su-Olson benchmark (Fig.~\ref{fig:SuOlsonBasic}) the new {\em discontinuous asymptotic $P_1$ approximation} yields better results than all other approximations. 
More also, in the Back experiment conditions, it finds the breakout time to be very close to exact simulations in 1D, better than the classic approximations (such as diffusion, $P_1$ or flux-limited diffusion). More complicated 2D exact simulations are still required, in order to complete the understanding of this experiment. These simulations will provide better estimations for the energy wall losses, and for the importance of the gold ablation. 

\appendix
\section{Numerical scheme for solving the boundary temperature equations}
\label{Appendix}

The square value of the total energy that is contained inside the foam can be calculated via Eq.~\ref{EHR}
and Eq.~\ref{HRXF}:
\begin{equation}
E^2(t)=f^2\rho^{2(1-\mu)}x_{F}^2(t)H^{2\varepsilon}(t)(1-\varepsilon)^2=f^2\rho^{2(1-\mu)}(2+\varepsilon)(1-\varepsilon)CH^{\varepsilon}(t)\int_0^t H(t')dt'
   \label{EHRSquare} 
\end{equation}
$\int_0^t H(t')dt'$ is separated to: $\int_0^t H(t')dt'=\int_0^{t-dt'} H(t')dt'+\int_{t-dt'}^t H(t')dt'$. We then define:
\begin{subequations}
\begin{equation}
Z_1=f^2\rho^{2(1-\mu)}(2+\varepsilon)(1-\varepsilon)C
\end{equation}
\begin{equation}
Z_2=Z_1\int_0^{t-dt'} H(t')dt'
   \label{Zconst} 
\end{equation}
\end{subequations}
Now Eq.~\ref{EHRSquare} can be rewritten as:
\begin{equation}
E^2(0,t)=\left(Z_2+Z_1\int_{t-dt'}^t H(t')dt'\right)H^{\varepsilon}(t)
   \label{EHRSquare2} 
\end{equation}
The algorithm's numerical step is as follows:
\begin{itemize}
\item {The surface temperature $T_S(t-dt)$ is taken from the last time step, solving for the flux using Eq.~\ref{MarshakBC}.}
\item{The new energy is calculated (The time integral over the flux), for Eq.~\ref{EHRSquare2}.}
\item{We solve for the new $H(t)=T_S^{4+\alpha}(t)$ using Eq.~\ref{EHRSquare2} numerically, and thus, finding the new $T_S(t)$.}
\item{Using the new $T_S$(t), we solve for the heat front $x_F(t)$ using Eq.~\ref{HRXF}.}
\end{itemize}
We repeat this step iteratively. 

\begin{acknowledgments}
We acknowledge the support of the PAZY Foundation under Grant No. 61139927. The authors thank Roee Kirschenzweig for using an IMC code, Roy Perry for using a $S_N$ code for radiative problems, and Guy Malamud and the anonymous referees for their valuable comments. Special Thanks to Mordecai (Mordy) Rosen from LLNL for the valuable conversations regarding the three different radiation temperatures and the wall-albedo.
\end{acknowledgments}

\bibliographystyle{apsrev4-1}

\end{document}